\documentclass[aps,prl,groupedaddress,twocolumn,showpacs]{revtex4}
\usepackage{graphicx}
\begin{document}
\title{Rabi flopping between ground and Rydberg states with dipole-dipole atomic interactions}
\author{T. A. Johnson, E. Urban, T. Henage, L. Isenhower, D. D. Yavuz, T. G. Walker, and M. Saffman}
\affiliation{Department of Physics, University of Wisconsin, 1150 University Avenue,
Madison, WI 53706}

\date{\today}

\begin{abstract}
We demonstrate Rabi flopping of small numbers  of $\rm{^{87}Rb}$ atoms 
between ground and Rydberg states with $n\le 43$. Coherent population oscillations are observed for  single atom flopping, while the presence of two or more atoms decoheres the oscillations. We show that these observations are consistent with van der Waals interactions of Rydberg atoms. 
\end{abstract}

\maketitle


Atoms in highly excited Rydberg states with principal quantum number $n>>1$ have  very large dipole moments which scale as $d\sim q a_0 n^2$, with $q$ the electron charge and $a_0$ the Bohr radius. Two such Rydberg atoms can be strongly coupled via a dipole-dipole interaction. It was recognized in recent years that the large interaction strength can potentially be used for fast quantum gates between qubits stored in stable ground states of neutral atoms\cite{ref.jaksch2000}.
When several atoms are sufficiently close together the presence of a single excited atom can cause a shift in the energy of all other atoms which is large enough to prevent resonant excitation of more than one atom in a sample. This ``dipole blockade" mechanism has the potential for creating strongly coupled ensembles containing moderate numbers of atoms. Such ensembles can be used for gates\cite{ref.lukin2001}, as well as several other quantum information tasks including state preparation\cite{ref.sw1}, fast measurement protocols\cite{ref.sw2}, and collective encoding of multi-qubit 
registers\cite{ref.brion}.  

A number of recent experiments have revealed signatures of ``dipole blockade" by showing that the probability of multiple excitation of Rydberg atoms is suppressed at high $n$ \cite{ref.suppression}. However, none of the experiments to date have demonstrated blockade at the level of a single atomic excitation which is crucial for applications to quantum information processing. In order to be useful for quantum gates it is also necessary to be able to coherently excite and de-excite a Rydberg state so that the atom is available for further processing.  In this letter we demonstrate important steps towards the goal of a fast neutral atom Rydberg gate. We start by preparing single atom states in micron sized optical traps and  observe coherent Rabi oscillations between ground and Rydberg states with $n\le 43$ at rates as high as $\Omega_R/2\pi = 0.5 ~\rm MHz.$ We then show that the presence of  two or more atoms in the trap causes dephasing of the Rabi oscillations. Comparison with theoretical calculations of the strength of the Rydberg van der Waals interactions\cite{ref.wstheory}, confirms that our observations are consistent with the presence of Rydberg interactions. 
 
The experiment starts by loading a far-off-resonance optical trap (FORT)
from a $^{87}$Rb vapor cell  magneto-optical trap (MOT) as described in our recent letter\cite{ref.yavuz2006}. 
For the experiments reported here, between 1 and 10 atoms are loaded into a 10~mK deep FORT (570~mW of 1030~nm light focused to a  $1/e^2$ intensity radius waist  of $w=2.7~\mu\rm m)$. 
The radial and axial oscillation frequencies are 130 and 12~kHz.
The average number of atoms is controlled by varying the amount of time for which the MOT and FORT lasers are simultaneously on from 25 - 400~ms. 
  Atom temperatures in the FORT are measured by performing a drop and recapture measurement, and comparing the probability of recapture
with numerical calculations. We consistently find temperatures from $5-10~\%$ of the FORT depth, which corresponds to $0.5-1~\rm mK$ for our typical parameters. These temperatures are much higher than Doppler cooling temperatures for $^{87}$Rb, which we attribute to  degradation of the laser cooling by  large FORT induced differential Stark shifts  of the
$5s_{1/2}$ and $5p_{3/2}$ levels. With a conservative estimate of  $1~\rm mK$ for the temperature, the spatial distribution of the atoms is   quasi one-dimensional 
with standard deviations of $\sigma_z=0.43~\mu\rm m$ and $ \sigma_x = 3.9~\mu\rm m.$ 

\begin{figure}[!t]
\includegraphics[width=7.5cm]{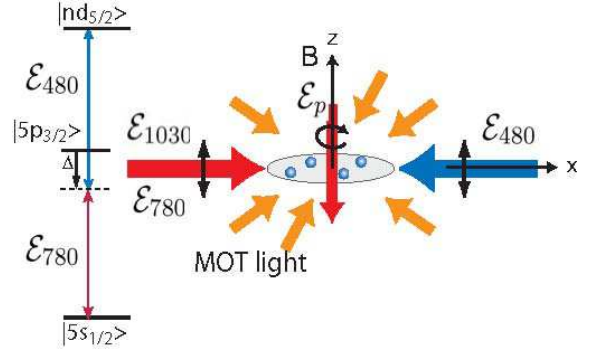}
\caption{\label{fig.experiment}(color online) Experimental arrangement, see text for details.}
\end{figure}

\begin{figure}[!t]
\includegraphics[width=7.5cm]{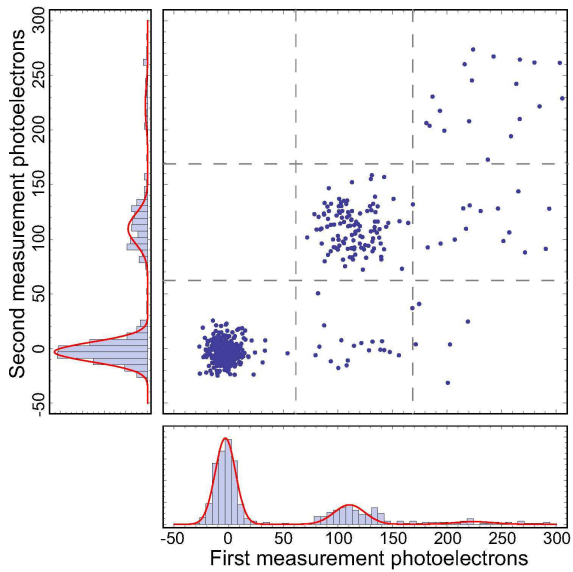}
\caption{\label{fig.readout}(color online) Correlation between first and second measurement distributions, 
without Rydberg excitation between measurements. The dashed lines show cuts for selecting single atom states. }
\end{figure}

We prepare single atom states in the FORT  despite Poissonian loading statistics by using a two-measurement sequence.  After the MOT to FORT loading period, we conduct a first measurement of the number of atoms in the FORT by scattering MOT light (3 pairs of counterpropagating beams) detuned by $-5~\rm MHz$ with respect to the cycling transition while chopping the FORT on and off at  rates between  $0.5 - 2.0 \times 10^6~\rm s^{-1}.$  The probing MOT light is chopped out of phase with the FORT, eliminating the need for tuning to the Stark shifted atomic resonance.   Scattered photons are collected with a fast lens (NA=0.4)  and focused onto a cooled electron-multiplying CCD camera. We estimate our detection efficiency including finite solid angle, optical losses, and camera quantum efficiency to be about 2.7\%. We observe single atom photoelectron rates of about $10^4~\rm s^{-1}$ (here time is the probing time with the MOT beams on, with the total  measurement time about $2.5$ times longer due to chopping) which gives the histogram shown on the abscissa of Fig. \ref{fig.readout} for a $12 ~\rm ms$ (probing time) first measurement.  
This is much shorter than the background gas limited $1/e$ FORT lifetime of about $3~\rm s.$
We verify the reliability of preselecting single atom states  by performing  a second measurement,  shown on the ordinate. We see that despite some loss of atoms during the first measurement, single atom states can be prepared with about 85\% probability, with a 15\% admixture of zero atom states. 
Note that the reliability of selecting states with two or more atoms is much worse.   We believe that this is due to light assisted collisions causing rapid loss out of the FORT during the first measurement \cite{ref.wieman1989}.

We excite Rydberg states using two-photon transitions with 780 and 480 nm lasers as shown in Fig. \ref{fig.experiment}. The Rydberg beams ${\mathcal E}_{780}, {\mathcal E}_{480}$ are focused to waists of $w\simeq 10~\mu\rm m$ and spatially overlapped with the FORT. These beams are generated by locking a 780 nm laser and a 960 nm laser to different longitudinal modes of the same 
stable reference  cavity with finesse ${\mathcal F}\sim~120,000$ and  linewidth $\sim 4~\rm kHz$. The cavity is constructed of ultralow expansion   glass
and is placed inside a temperature stabilized vacuum can. We obtain long term instability of a few hundred kHz, 
and short term instabilities of both lasers relative to the cavity line of a few hundred Hz at $<10~\mu\rm s$ averaging time. The 960 nm light is then amplified and frequency doubled to create the 480 nm Rydberg excitation light.  Acousto-optic modulator (AOM) based noise eaters are used as necessary to reduce amplitude fluctuations to a few percent.   The frequencies of both lasers are then shifted with AOM's to match the desired Rydberg level.

\begin{figure}[!t]
\includegraphics[width=8.5cm]{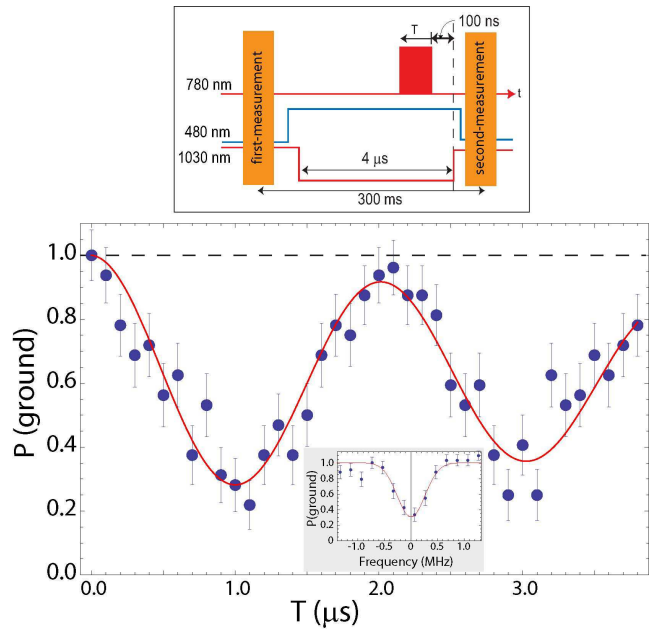}
\caption{\label{fig.43d oneatom}(color online) Rabi flopping experiment to $43d_{5/2}$
with $P_{780}=1.85~\mu\rm W, P_{480}= 10.7~\rm m W$, and $\Delta/2\pi = -3.4~\rm GHz.$
 Each data point is the average of 40 pre-selected single atom experiments, with the bars showing $\pm$ 1 standard deviation.  The inset shows spectroscopy of the resonance obtained by scanning the frequency of the 780 nm light.  }
\end{figure}

After preparing a single atom state the probability of transition to a Rydberg level is measured as a function of the pulse length of ${\mathcal E}_{780}, {\mathcal E}_{480}.$ If the atoms are not prepared in a single ground state Zeeman level they will be coupled to a superposition of different Rydberg Zeeman levels. In the presence of background magnetic fields these levels have different Zeeman shifts which decohere the coherent population oscillations we are interested in. We therefore start by optically pumping into the  $|f=2, m_f=2\rangle$ Zeeman state  using $\sigma_+$ polarized light (${\mathcal E}_p$ in Fig. \ref{fig.experiment}) near resonant to the $|5s_{1/2}f=2\rangle\leftrightarrow| 5p_{3/2}f'=3\rangle$ transition with a $B=10^{-3}~\rm T$ bias field. The $\hat z$ polarized excitation light only couples this state to the $|nd_{5/2},m_j=1/2\rangle$ Zeeman state. We have verified the expected shift of $-5.6~\rm  MHz$ of the Rydberg excitation frequency in the presence of the bias field by performing spectroscopy  with and without the field on. 

\begin{figure}[!t]
\includegraphics[width=8.cm]{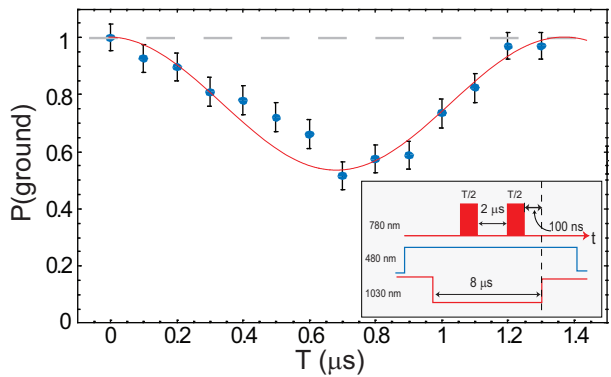}
\caption{\label{fig.doublepulse}(color online) Rabi flopping experiment with single atoms to $28d_{5/2}$ using a double pulse sequence.
The excitation parameters were the same as in Fig. \ref{fig.43d oneatom} except that $P_{780}=3.3~\mu\rm W, P_{480}= 9.0~\rm m W$, and $\Delta/2\pi = -3.8~\rm GHz.$
The solid curve is a theoretical calculation assuming $\Omega_R/2\pi=0.7~\rm MHz$ and a detuning in the 
$2~\mu\rm s$ gap of $0.53~\rm MHz.$
 The inset shows the timing sequence.  }
\end{figure}

In order to measure the probability of Rydberg excitation we use the fact that the calculated photoionization rate of the $d$ states due to the FORT light is large compared to the radiative decay rate back to the ground state\cite{ref.swpra2005}. The timing sequence for Rabi flopping is shown in 
Fig. \ref{fig.43d oneatom}. We turn off the FORT light for a fixed length of time which is long enough to perform the Rydberg excitation yet short enough that we do not lose the atom in the absence of a Rydberg pulse. 
We then perform a Rydberg pulse of variable length $T$, after which we restore the FORT light. Photoionization by the FORT light thus performs a projective measurement of the atomic state, and after 100 ms with the FORT on, the ground state population is measured using MOT light. The resulting data points for the ground state probability are then normalized to $1.0$  at $T=0$ to correct for a 10-20\% loss rate due to the 
single atom selection measurement, motional losses during the FORT drop period, and FORT loss due to background collisions.  A curve fit to the data of Fig. \ref{fig.43d oneatom} with the function $(1-a)+a e^{-\frac{t}{\tau}}\cos(\Omega_R t)$ gives a Rabi frequency of $\Omega_R=2\pi\times0.49~\rm MHz$, whereas our theoretical value with no adjustable parameters 
is\cite{rabicalc} $\Omega_R=\Omega_{780}\Omega_{480}/2\Delta=2\pi\times 0.55~\rm MHz$.  We attribute the approximately 11\% lower experimental value to spatial misalignment, and a small fraction of the Rydberg light being present in servo sidebands from the laser locks. The fit also gives a decay time of $\tau=8~\mu\rm s$ which is consistent with Doppler averaged numerical calculations.  The data show that the atom is returned to the ground state with better than 90 \% probability and that Rydberg state excitation is achieved with $70-80 \%$ probability. The lack of perfect Rydberg excitation is due to several factors  which we estimate as 
Doppler broadening of the excitation($\sim 10-20\%)$, imperfect optical pumping ($\sim 5\%)$,  and imperfect detection efficiency ($\sim 4\%)$\cite{ratenote}.

A quantum gate protocol\cite{ref.jaksch2000} requires that we leave an atom in one site in a Rydberg level while
a conditional Rydberg excitation is performed at a neighboring site. With our observed $\pi$ pulse time of $\sim 1~\mu\rm s$ this requires staying in the Rydberg level for at least $2~\mu\rm s$. We have verified that coherence can be maintained over this time by performing a double pulse experiment as shown in Fig. \ref{fig.doublepulse}.   For this experiment the maximum probability of Rydberg excitation is only about 50\%.
This is because during the $2~\mu\rm s$ free evolution between the pulses the 780 nm Rydberg laser induced light shifts are no longer present which results in a free evolution for $2~\mu\rm s$ at finite detuning.
We have fit the measured data by assuming a detuning of  $ 0.53~\rm MHz$. This is consistent with  the 
780 nm beam induced light shift of the ground state which is $\Omega_{780}^2/4\Delta=2\pi\times 0.58~\rm MHz$ for our estimated experimental parameters.

\begin{figure}[!t]
\includegraphics[width=8.5cm]{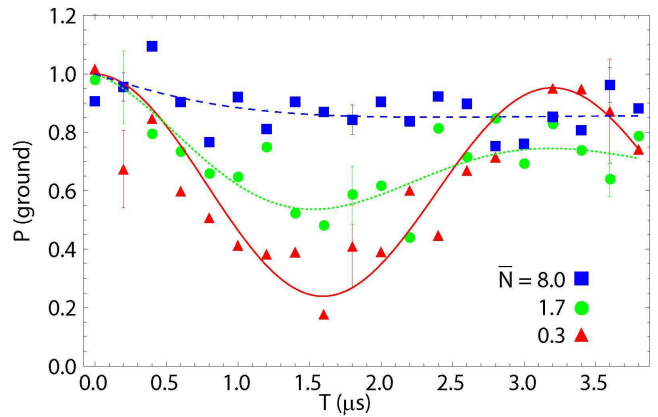}
\caption{\label{fig.43dmanyatoms}(color online) Rabi flopping experiment to $43d_{5/2}$ with different atom numbers.
 Each data point is the average of 150 measurements, with the bars showing $\pm$ 1 standard deviation.}
\end{figure}

The above Rabi flopping curves change dramatically when more than one atom is present in the FORT. 
Figure \ref{fig.43dmanyatoms} shows the results with average atom numbers of $\bar N = 0.3, 1.7,$ and $8$.
These data were obtained by simply adjusting the loading time for a desired $\bar N$ (the actual distribution is close to Poissonian), and then applying varying length  Rydberg pulses, without first using the single atom selection procedure of Fig. \ref{fig.readout}. We see that at $\bar N=1.7$ the visibility of the oscillations is strongly reduced, and at $\bar N=8$ there is essentially no oscillation left.  
We emphasize that Rabi flopping with high visibility and good coherence between ground state levels 
 is observed with as many as 10 atoms in the FORT\cite{ref.yavuz2006}. 

In order to explain these observations we must account for Rydberg interactions. In the absence of an external electric field $^{87}$Rb atoms excited to the $43d_{5/2}$ state experience a F\"orster interaction due to the near resonance of the process $43d_{5/2}+43d_{5/2}\rightarrow 45 p_{3/2} + 41 f_{5/2,7/2}.$ The corresponding energy defect 
is $ \hbar \delta=U(45p_{3/2})+U(41f_{5/2,7/2})-2U(43d_{5/2}) .$ 
Using recent measurements of the Rb quantum defects\cite{ref.gallagher}, we find $\delta/2\pi = -6.0, -8.3~\rm MHz$ for the $f_{5/2}$ and $f_{7/2}$ states respectively. This small energy defect naively implies a long range $1/R^3$ interaction, with $R$ the atomic separation. However, due to Zeeman degeneracy of the Rydberg levels, there are linear superpositions of two-atom $d_{5/2}$ states which are excited by the Rydberg lasers, yet have small dipole-dipole 
interactions\cite{ref.wszeroes}. We have developed a full theory of this situation which will be published elsewhere\cite{ref.wstheory}. For the present discussion we note that in the limit where the atoms have a relatively large separation we obtain a van der Waals interaction of the form $V_{\rm dd}=C_6 D_\phi/R^6,$ with $C_6\sim d^4/\hbar \delta $, and $D_\phi$ 
an eigenvalue. The eigenvalues depend on $M$, the projection of the two-atom angular momentum on the interatomic axis, and for the $|d_{5/2}\rangle$ states very small values of $D_\phi$ occur for $M=0.$
In  our experimental  geometry we excite  states with $M_z=1$ along $\hat z,$ but in a quasi one-dimensional FORT large values of $R$ occur for atom pairs aligned along $\hat x.$ States with $M_z=1$ are linear superpositions of states with $-5\le M_x\le 5$, and the states with $M_x=0$ overlap with eigenstates for which  $D_\phi$ is very small.

\begin{figure}[!t]
\includegraphics[width=8.5cm]{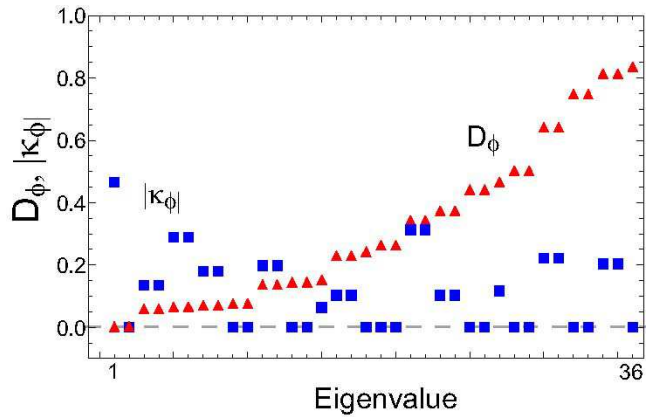}
\caption{\label{fig.overlap}(color online) Eigenvalues (triangles) and overlap factors (squares) for the two-atom van der Waals eigenstates with $-5\le M_x\le 5$. }
\end{figure}

 Figure \ref{fig.overlap} shows the overlap coefficients $\kappa_\phi=\langle\phi|M_z=1\rangle$,  together with the eigenvalues $D_\phi$ for the $(2j+1)^2=36$ eigenvectors. We see that 
there is a strong overlap with the state with the smallest eigenvalue, $D_{\phi,\rm min }=0.0024.$ 
For this interaction   $|43d_{5/2}\rangle $ has $C_6 = 450~\rm GHz~ \mu m^6 $ and at a characteristic two-atom separation in the FORT of $\sqrt{\langle R^2 \rangle} =\sqrt2 \sigma_x =7.8~\mu\rm m $ the interaction strength due to the smallest eigenvalue is only  $4.8~\rm kHz$ which is much smaller than the excitation Rabi frequency. These ``F\"orster zero" states prevent an effective blockade despite the large value of $C_6.$ At the same time 
there is strong coupling to states with $D_{\phi}= 0.81$ which give a characteristic interaction strength of 
$1.6~\rm MHz$. We thus have a situation where two-atom excitation occurs  because of F\"orster zero states, and at the same time other eigenstates lead to interaction strengths which are larger than our excitation Rabi frequency. In this regime the two-atom states experience a strong dephasing from two-atom shifts, which results in the washing out of Rabi oscillations seen in Fig. \ref{fig.43dmanyatoms}. This decoherence is also expected due to the finite extent of the FORT and consequent variations in $R$, but is further accentuated by coupling to both small and large eigenvalues.    In the limits of very weak interactions, or very strong interactions giving an effective blockade, we do not expect rapid dephasing in the multiple atom regime, which is a signature of a range of coupling strengths.

In conclusion we have observed coherent Rabi oscillations between ground and Rydberg states which is an important step towards demonstration of a neutral atom Rydberg gate. We have shown how van der Waals interactions lead to dephasing of the oscillations when several atoms are present, and elucidated the 
role of F\"orster zero states in the dephasing. Future work will explore alternative interaction geometries that do not couple to small eigenvalues as a means of demonstrating dipole blockade.

This work was supported by the NSF and the Disruptive Technology Office. We are grateful to Pasad Kulatunga and Marie Delaney for experimental contributions at an early stage of this work.


\begin{thebibliography}{99}
\bibitem{rabicalc}
For $\hat z$ polarized light $\Omega_{780}=\frac{\sqrt2}{3}\frac{q{\mathcal E}_{780}}{\hbar}
R_{5_{s_{1/2}}}^{5_{p_{3/2}}}$ and 
$\Omega_{480}=\frac{\sqrt6}{5}\frac{q{\mathcal E}_{480}}{\hbar}R_{5_{p_{3/2}}}^{43_{d_{5/2}}}$.
Using quantum defect wavefunctions we find  
$R_{5_{s_{1/2}}}^{5_{p_{3/2}}}=5.14a_0, 
R_{5_{p_{3/2}}}^{43_{d_{5/2}}}= 0.029a_0.$
\bibitem{ratenote}
The ratio of room temperature radiative transitions to photoionization rates is about $(1.3\times 10^4~{\rm s}^{-1})/(3.1\times 10^5 ~{\rm s}^{-1})$ which implies a $4\%$ reduction 
in our measured excitation probability relative to the true value.
\end{thebibliography}
\end{document}